# A Proposal for a Solution to the Accommodation-Vergence Mismatch Problem in 3D Displays

*Dal-Young Kim*
Department of Optometry, Seoul Tech, 232 Gongneung-ro, Nowon-gu, Seoul 139-743, Korea
Tel.:82-2-970-6229, E-mail: dykim@seoultech.ac.kr

### Abstract

*Here is suggested a solution to the accommodation-vergence mismatch problem in 3D stereoscopic displays. It can be achieved by compensating the mismatched focal length with refractive power of adjustable-focus 3D glasses. The compensation would make vergence of human eyes match with position of virtual stereoscopic motion pictures, reducing visual fatigue.*

### 1. Introduction

The accommodation-vergence mismatch (AVM) has been considered as a cause of the visual fatigue induced by watching three-dimensional (3D) stereoscopic displays,[1-3] though some recent researches suggested that it was not an essential cause.[4,5] Owing to rapid growth of the 3D TV and cinema industry, much attention is paid to it. Various ideas have been proposed to solve the AVM problem, and most of solutions are belonged to one of two categories, that is, the wave front reconstructing displays or the volumetric displays.[6] Differently from those previous solutions, we would propose a solution based on physiological optics and adjustable-focus 3D glasses.

### 2. The Accommodation-Vergence Mismatch

The AVM is a mismatch between focal length (adjusted by accommodation) of human eye and binocular vergence (convergence or divergence) angle when a viewer watches the 3D display or head-mounted display. Fig. 1 (a) depicts the viewer is watching a conventional two-dimensional (2D) display. Here $N_1$ and $N_2$ are the nodal points of left and right eyeballs. Visual axes of the viewer's both eyes converge to a fixation point (FP) on the 2D display, and the vergence angle is defined as $\theta_1$. The focal lengths of both eyes are adjusted to $N_1$-FP and $N_2$-FP for left and right eye, respectively. On the other hand in Fig. 1 (b), the viewer is watching a (2-view type) 3D stereoscopic display. Every frame of displayed motion pictures is divided into two pictures for stereoscopic vision. One picture is incident from $FP_1$ to left eye while the other picture is incident from $FP_2$ to right eye, inducing binocular disparity. The human stereopsis produces a virtual 3D image at V point to which the visual axes of viewer's two eyes converge. The convergence angle changes from $\theta_1$ to $\theta_2$. However, the focal lengths of left and right eyes are neither $N_1$-V nor $N_2$-V, but $N_1$-$FP_1$ and $N_2$-$FP_2$, respectively. It is because the 3D image at V point is not a real one but real pictures are still on the surface of 3D display.

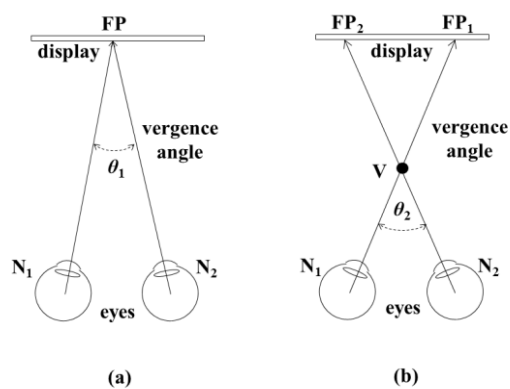

**Fig. 1.** Relation between the fixation points, vergence angle and images when a viewer watches (a) a 2D display or (b) a 2-view type 3D display.

It is well known that the vergence and accommodation are not independent but highly correlated with each other.[7] The focal lengths have to shorten with wide vergence angle when the viewer watches a near point, while they have to lengthen with narrow vergence angle when the viewer watches a far point. This coupling of vergence and accommodation is so strong that the viewer can feel discomfort if it is broken. Long focal lengths with wide vergence angle or short focal lengths with narrow vergence angle are quite unusual. The situation of Fig. 1 (a) is natural and comfortable for the viewer, but that of Fig. 1 (b) is discomfort due to abnormal long focal lengths with wide vergence angle. This is the AVM.

### 3. Physiological-Optical Compensation

In the view point of the physiological optics, the most simple but efficient solution to the AVM problem is to compensate the mismatched accommodation (focal length) by refractive power of lenses.

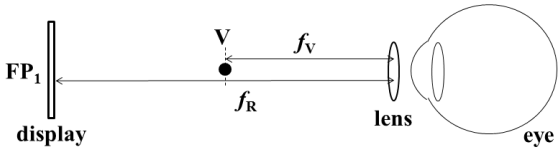

**Fig. 2. Focal lengths of eye and a lens of 3D glasses.**

Fig. 2 describes relations between the focal lengths of the left eye and a spectacles-lens. Here let me introduce the concept of refractive power ***D***, which is defined as an inverse of focal length. It is well known that total refractive power ***D*$_R$** of a certain ophthalmic-optical system is approximately a sum of refractive powers of the eye and the lens. Taking the refractive power of lens into account, ***D*$_R$** in Fig. 2 is given as

$$D_R \approx D_V + D_C \qquad (1).$$

Here, ***D*$_V$** and ***D*$_C$** are refractive powers of the viewer's eye and the lens. In order to watch the real image (FP$_1$) on the surface of 3D display, the focal length and refractive power of the total system must be N$_1$-FP$_1$ (*f*$_R$) and 1/*f*$_R$. As mentioned above, the viewer feels comfortable when the focal length and refractive power of eye are N$_1$-V (*f*$_V$) and 1/*f*$_V$. From these constraints, focal length (*f*$_C$) and refractive power (***D*$_C$**) of the lens required to satisfy Eq. (1) is calculated as;

$$D_C = D_R - D_V = \frac{1}{f_R} - \frac{1}{f_V} \equiv \frac{1}{f_C} \qquad (2).$$

If *f*$_C$ satisfies above Eq. (2), the viewer can watch comfortably the real image (FP$_1$) by total refractive power of *f*$_R$, with natural coupling of convergence to the V point and eye focal length of *f*$_V$. Eq. (2) is also valid in case of left eye, or when the virtual image V locates at the back of the 3D display, resulting in negative focal length of *f*$_C$ that means the lens is a concave one. This compensation by the refractive power of lenses is an old solution to the AVM problem, which was adopted by inventors of stereoscope even in the 1800s.[8]

## 4. A Solution the AVM Problem

At a glimpse, this compensation method looks impractical for 3D motion pictures because it requires continuous change of the focal lengths of lenses. However, recent invention of adjustable-focus eye-glasses[9] made the compensation method practical for the 3D motion pictures as well.

The adjustable-focus eye-glasses are composed of liquid crystal (LC) panels that electro-optically diffract lights like lenses. By changing external electric field applied to the LC, their focal lengths and refractive powers are continuously adjustable. They were primarily designed for presbyopia as an alternative to the progressive lenses.[9]

If this adjustable-focus LC panels are adopted in the lenses of 3D glasses, continuous change of the *f*$_C$ that is required for the compensation method will be available, so that the viewer will comfortably watch 3D motion pictures as discussed in section 3. This is the solution that we would propose.[10]

We have to know the position of V point to calculate the required *f*$_C$. It is not easy to estimate the depth perceived by binocular vision, due to its complicate mechanism.[11] However, geometrical position of the V point may be computationally calculated from the viewer's position and inter-pupillary distance, the distance between FP$_1$ and FP$_2$ in Fig. 1 (b), and so on. All of these factors can electronically be detected or can be input by the viewer.

In the above discussion about Fig. 2, for simplicity, we did neither consider interval between the lens and eyeball. This inter-vertex distance must also be considered for precise calculation of *f*$_C$ in Eq. (2).

## 5. Summary

A solution to avoid the AVM problem in 3D display was proposed. It is the compensation of AVM by the refractive power of adjustable-focus 3D glasses. To realize this scheme, further theoretical, instrumental, and clinical studies are needed.

**Acknowledgement**

This work was financially supported by LG Electronics.